\documentclass[a4paper,twoside]{article}

\usepackage{epsfig}
\usepackage{subcaption}
\usepackage{calc}
\usepackage{amssymb}
\usepackage{amstext}
\usepackage{amsmath}
\usepackage{amsthm}
\usepackage{multicol}
\usepackage{pslatex}
\usepackage[authoryear]{natbib}
\usepackage{algorithm2e}
\usepackage[bottom]{footmisc}

\usepackage{url}
\usepackage[hidelinks]{hyperref}
\usepackage{gensymb}
\usepackage{todonotes}

\usepackage{SCITEPRESS}     

\begin{document}

\title{Measuring Braking Behavior Using Vehicle Tracking and Camera-to-Satellite Homography Rectification}

\author{\authorname{
J.P. Fleischer\sup{1}\orcidAuthor{0000-0002-1102-1910}, Tanchanok Sirikanchittavon\sup{2}\orcidAuthor{0009-0004-9829-4584}, Chonlachart Jeenprasom\sup{2}\orcidAuthor{0009-0006-2543-4616}, Nooshin Yousefzadeh\sup{1}\orcidAuthor{0000-0003-1232-5972}, Sanjay Ranka\sup{1}\orcidAuthor{0000-0003-4886-1988} and Mohammed Hadi\sup{2}\orcidAuthor{0000-0003-2233-8283}}
\affiliation{\sup{1}Department of Computer and Information Science and Engineering, University of Florida, Gainesville, FL, USA}
\affiliation{\sup{2}Department of Civil and Environmental Engineering, Florida International University, Miami, FL, USA}
\email{jacquespfleischer@gmail.com}
}


\abstract{This paper presents an open-source software application for analyzing traffic camera footage, focusing on vehicle behavior and braking events at signalized urban highways. The core innovation is a robust ground-plane homography estimation that links fixed traffic camera views to satellite orthoimagery. This process rectifies the camera's oblique perspective, ensuring that pixel distances accurately represent real-world distances. This enables the acquisition of features such as vehicle trajectory, speed, deceleration, and braking severity without the need for camera calibration. The pipeline employs the MAGSAC++ estimator to build the homography, converting YOLO11 object detections into a rectified top-down coordinate system. All detection and trajectory data are stored in a ClickHouse database for subsequent analysis. A real-world case study at two signalized intersections in Key West, Florida, showcased the system's capabilities. Across two days of daytime footage, braking activity at the higher-volume intersection peaked around 4 PM at approximately 57.5 events per hour, while the second intersection peaked around 10 AM at roughly 15.5 events per hour. The spatial analysis revealed that most braking events initiated upstream, with mild and moderate braking mostly occurring 30–45+ meters away from the stop bar and severe braking distributed throughout, but particularly concentrated in lanes with higher interaction and merging activity. The findings highlight the significant potential of this centralized safety information system to support connected vehicles, facilitating proactive traffic management, crash mitigation, and data-driven roadway design and safety analysis.}

\onecolumn \maketitle \normalsize \setcounter{footnote}{0} \vfill

\section{Introduction}
\label{sec:introduction}

Performance measurement in real-time operations and offline analysis is critical to effective planning, design, operations, and management of the transportation systems. Performance measurement is also critical to providing information and guidance to connected vehicles to enhance the safety and mobility of these vehicles and the transportation system. Infrastructure-to-vehicle communications can be used for the transmission and notification of traffic measures as well as existing or predicted severe events, such as upstream traffic incidents or sudden stops. Such notifications offer a powerful and reliable source of information to distribute to connected vehicles and cooperative automated vehicles to improve performance.  

Hard-braking behavior is a significant surrogate indicator of crash risk, particularly at signalized intersections. A study comparing 4.5 years of rear-end crash data at eight signalized intersections with hard-braking event data found a strong correlation between hard-braking frequency and rear-end crashes using Spearman’s rank-order correlation~\citep{article}. Specifically, hard-braking events occurring more than 400 feet upstream of the stop bar were found to be strongly associated with rear-end crash occurrences.


Connected and automated vehicle (CAV) technologies have been shown to influence braking behavior at signalized intersections~\citep{Do_Saunier_2024}. For example, an evaluation of an Advanced Stop Assist System (ASAS) using a microscopic simulation~\citep{sengupta2026bigsumo} model demonstrated that Vehicle-to-Infrastructure (V2I) communication, providing advisory speed messages, could prevent hard braking at yellow or red signals. This system reduced hard-braking events by nearly 50\% at full market penetration, with sensitivity analyses suggesting that a 60\% penetration rate is necessary to minimize rear-end collisions and enhance safety at signalized intersections. Vehicle-to-vehicle communications can also communicate notifications of severe events, such as upstream traffic congestion or sudden stops~\citep{Lyu_Cheng_Zhu_Zhou_Xu_2019}. Integrating such notifications within traffic infrastructure, through V2I communications with devices like roadside units, can offer a broader and more centralized source of information for connected vehicles. The prompt communication of detected braking events is crucial for safety, enabling smart vehicles to respond quickly, thereby smoothing traffic flow and protecting road users.

Developing methods to detect braking events from conventional traffic cameras, rather than relying on V2X (Vehicle-to-Everything) connected vehicle data or advanced mapping technologies, addresses a significant research gap. Utilizing connected vehicle data for this purpose faces challenges due to the lower market penetration and feasibility of retrieving this data. Current object localization techniques often require high-definition base maps, which are not widely available and can be costly. Some homography-based traffic analysis approaches convert angled camera views to top-down perspectives but do not allow trajectories to be overlaid on geographic information systems (GIS) layers for robust estimation. Other studies on density estimation use detailed maps to retrieve satellite images, often necessitating camera calibration. This paper presents a novel approach that overcomes these limitations by utilizing a GIS platform (ArcGIS) and publicly available satellite imagery, thereby eliminating the need for costly high-definition maps and extensive camera calibration.

We detail the steps undertaken to create this pipeline and illustrate the application's effectiveness through a real-world case study at two signalized intersection in Key West, Florida. 

This application employs an open-source pipeline designed with main contributions include:
\begin{enumerate}
\item Facilitate data collection and straightforward recognition of the highway facility configuration and geometry.

\item Detect and track vehicles using advanced machine learning-based computer vision models.

\item Effectively detect and store braking events within an organized database schema for future analysis.
    
\end{enumerate}

The analysis of data collected in this case study demonstrates the application's ability to identify clear temporal and spatial patterns in braking behavior that can be correlated with crash occurrence. 

The rest of this paper is organized as follows. Section 2 describes the locations and specifications of the applied case study. Section 3 presents the architecture of the developed software application and briefly introduces the video analysis and orthorectification algorithms. Section 4 presents the experimental results from applying the developed application within a case study. Section 5 discusses related and future works, providing context and outlining potential directions for further research. Finally, Section 6 ends the paper with main conclusions.

\section{Case Study Site Specifications}
Two intersections in Key West, Florida were used in a case study to demonstrate the application developed in this study. The first intersection is a T-intersection junction type between Overseas Highway (U.S. Route 1) and Roosevelt Boulevard (State Road A1A). The camera is roughly located at 24.56998\degree N, 81.75266\degree W and points at the southbound queue. The second intersection is White Street and Truman Avenue; its camera is roughly located at 24.55619\degree N, 81.79134\degree W and points at the eastbound queue. The footage for both intersections was taken from 7:00 AM to 7:00 PM on February 13th and 14th, 2025. 

\section{System Architecture}

\subsection{System Overview}

The developed application ingests video taken from on-site rectilinear traffic cameras. After a one-time user supplies information regarding the intersection configuration, including performing feature matching (identifying key points in one photo and selecting the same points in another photo) and labelling stop bars, the system can generate metrics of vehicle trajectories and braking. These metrics are made possible via a server equipped with a GPU (graphics processing unit).

\subsection{Hardware}


The hardware used in this study includes roadside EZVIZ cameras used by a consultant to the Florida Department of Transportation to record videos for later manual processing (such as recording the number of vehicles for each turn movement at an intersection). The computer used to perform the processing in this study contains a NVIDIA RTX A6000 GPU, which is sufficiently capable of processing hours of traffic footage and performing object tracking as described later in this paper.


\subsection{Software}

The utilized pipeline in developing the application of this study consists of three software components, meant to perform pairing correspondence, object tracking, and braking analysis, as described in the following subsections.

\subsubsection{RoadPairer: Traffic Camera-to-Satellite Correspondence}
\label{sec:correspondence}

The \textit{RoadPairer}, as defined in this paper,  refers to a developed utility in this research. The utility allows a user to identify the configuration of a new intersection for analysis in the pipeline. To accomplish this, the process uses satellite imagery to pair with traffic camera frames. Satellite images for Monroe County, Florida, are available through its public ArcGIS REST service~\citep{MonroeCounty_Orthos2024}. The only input required to download a satellite image is two sets of longitude and latitude that correspond to the location site (\texttt{minLon, minLat, maxLon, maxLat}). The images are in GeoTIFF format, utilizing CRS (coordinate reference system) EPSG:6438 (established by the European Petroleum Survey Group). The georeferenced raster allows for standardizing the ground distance per pixel. 

With a satellite image and a native frame from the traffic camera, the user is able to create points across each camera view, making one-to-one correspondences. For example, to make the correspondence, the user clicks one point on a corner of the street median, then clicks another point on that same median corner in the other view. This feature matching should continue for as many features as possible, ideally spread out across the environment. This point correspondence is demonstrated in~(\ref{eq:correspondence}).


\begin{equation}
\label{eq:correspondence}
(u_i, v_i)_{\text{cam px}} \;\longleftrightarrow\; (x_i, y_i)_{\text{ortho px}}
\end{equation}

\begin{figure}[t]
    \centering

    \includegraphics[width=0.8\linewidth]{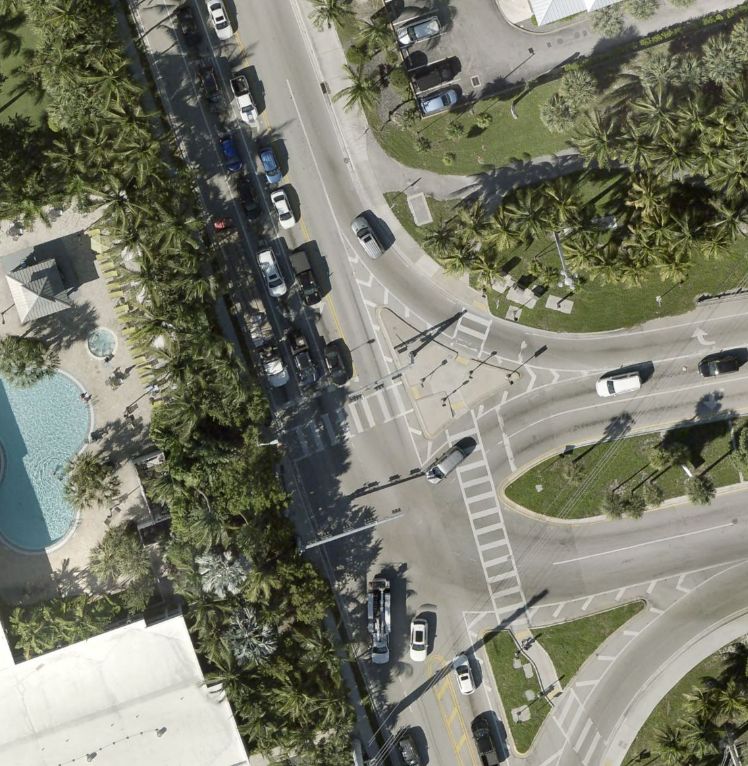}
    \caption{Satellite image of the Overseas Hwy.\ and Roosevelt Blvd.\ intersection from Monroe County.}
    \label{fig:orthozoom}

    \vspace{0.6em}

    \includegraphics[width=1.2\linewidth]{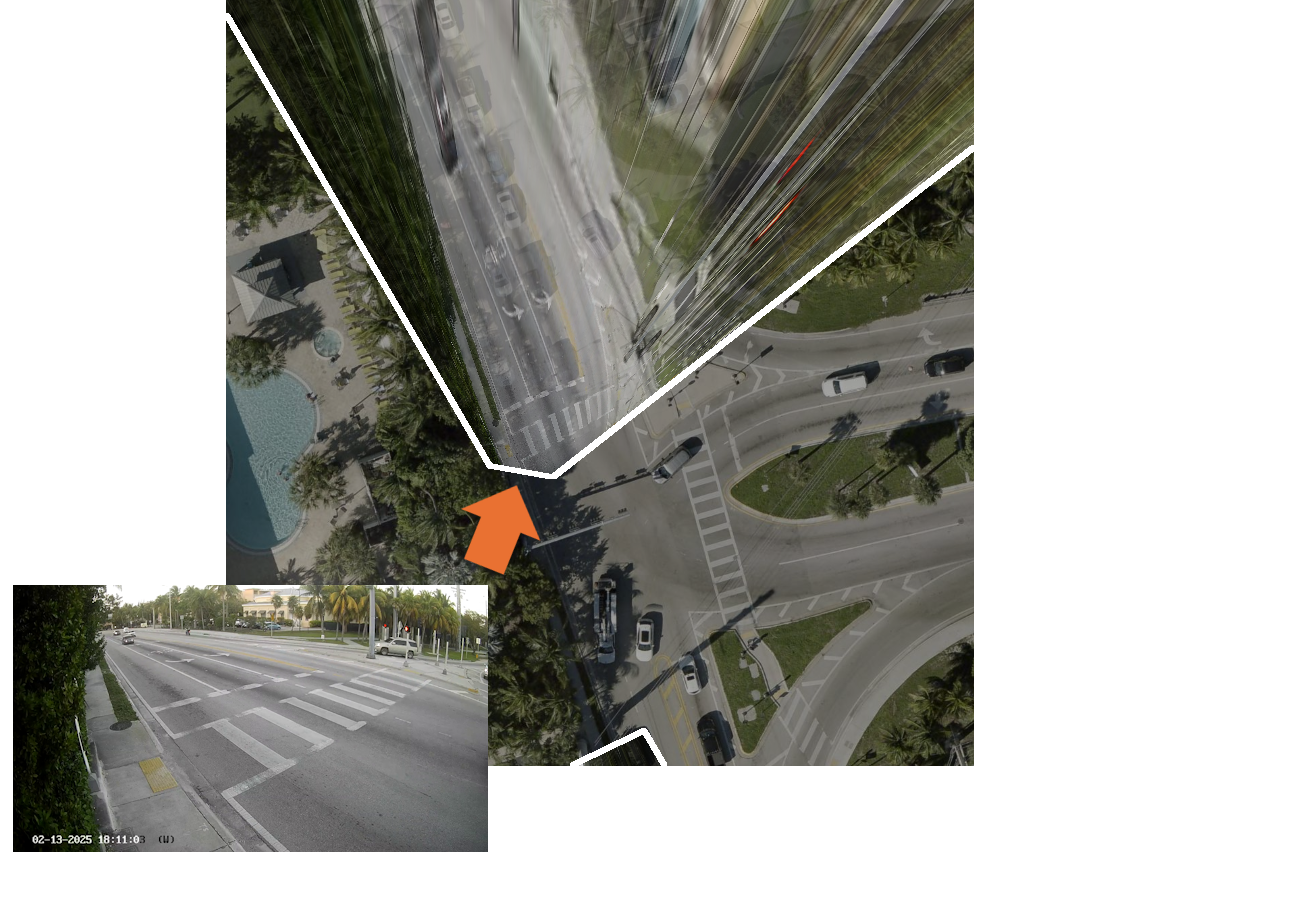}
    \caption{Projected traffic camera view overlaid on the satellite orthophoto.}
    \label{fig:camera_overlay}
\end{figure}

Once around 10 to 20 point pairs are created, a robust homography can be built. This homography can map points in the camera space to points in the satellite (orthography) space, as depicted in Figure~\ref{fig:camera_overlay}. The homography is built using the MAGSAC++, a state-of-the-art machine vision algorithm and robust estimator~\citep{barath2019magsacfastreliableaccurate} in the form of a $3\times3$ NumPy~\citep{harris2020array} matrix, which is depicted in Equation~\ref{eq:numpymatrix}. These values serve different functions within the homography; the upper-left $2\times2$ submatrix is used for the linear transform, rotating, scaling, and shearing functions. The remaining matrix terms are responsible for translation and projective effects.

\begin{equation}
\label{eq:numpymatrix}
\mathbf{H} =
\begin{bmatrix}
h_{11} & h_{12} & h_{13} \\
h_{21} & h_{22} & h_{23} \\
h_{31} & h_{32} & h_{33}
\end{bmatrix}
\end{equation}

Equation~\ref{eq:orthocoordinate} is used to compute the trajectory point coordinates ($x_i, y_i$) in the satellite (orthography) space for later analysis. $w_i$ in Equation~\ref{eq:orthocoordinate} denotes the third homogeneous coordinate obtained from the matrix-vector product after applying the homography; this third coordinate enables projective transformations that would not otherwise be possible with a two-dimensional matrix.

\begin{equation}
\label{eq:orthocoordinate}
\begin{bmatrix}
x_i \\
y_i
\end{bmatrix}
=
\frac{1}{w_i}
\mathbf{H}
\begin{bmatrix}
u_i \\
v_i \\
1
\end{bmatrix},
\qquad
w_i \neq 0
\end{equation}

The developed software application also contains functionality to mark the street median yellow line and the stop bar within the intersection. This enables the determination of vehicle distance from the stop bar at any point in time, as well as the differentiation between the two directions of travel.

\begin{figure*}[tb]
    \centering
    \includegraphics[width=0.85\linewidth]{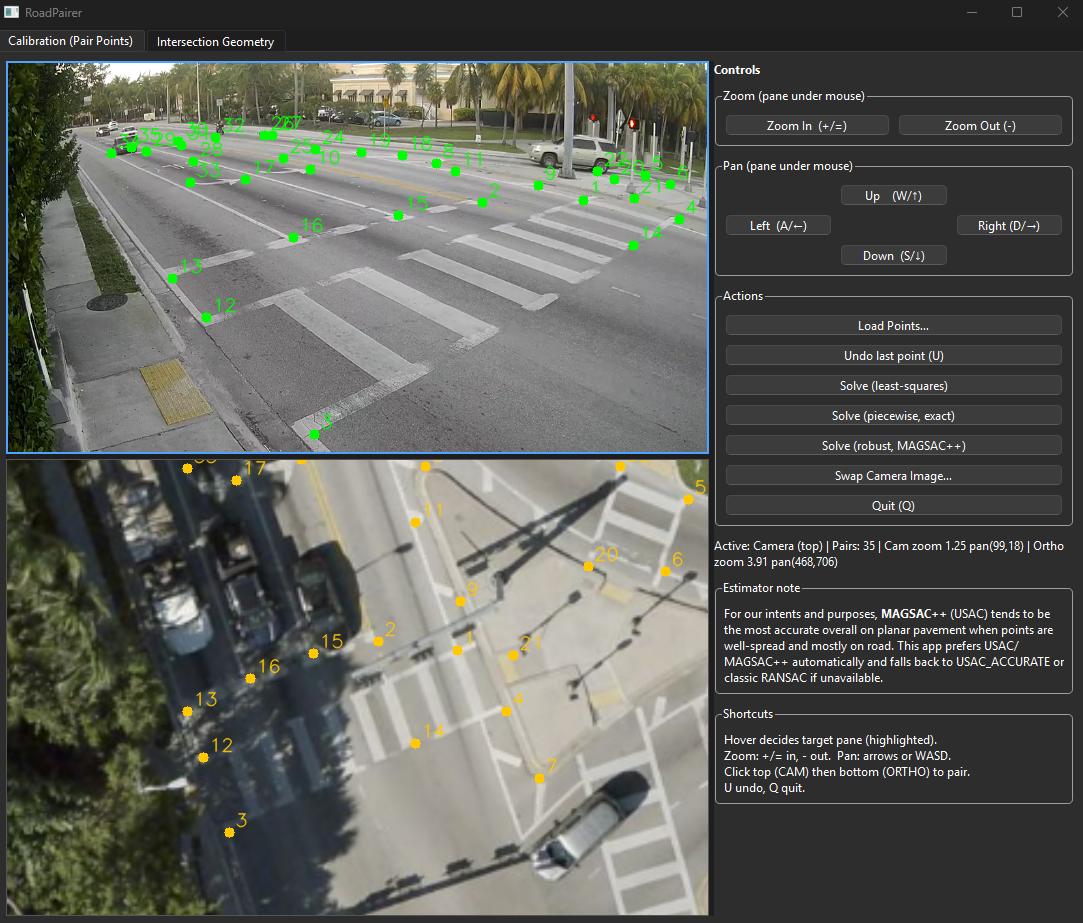}
    \caption{A screenshot of the RoadPairer GUI mapping points between a native traffic camera image and a satellite image.}
    \label{fig:roadpairergui}
\end{figure*}

The \textit{RoadPairer} utility, which generates the homography, is shown in Figure~\ref{fig:roadpairergui}. \textit{RoadPairer} allows panning, zooming, saving of points to retrieve at a later time, and other related features. The software utility is made open-source\footnote{\url{https://github.com/jpfleischer/rectangle-traffic-camera/}} and is automatically compiled for Windows in \texttt{.exe} format, so no installation is required.

\begin{figure*}[tb]
    \centering
    \includegraphics[width=0.98\linewidth]{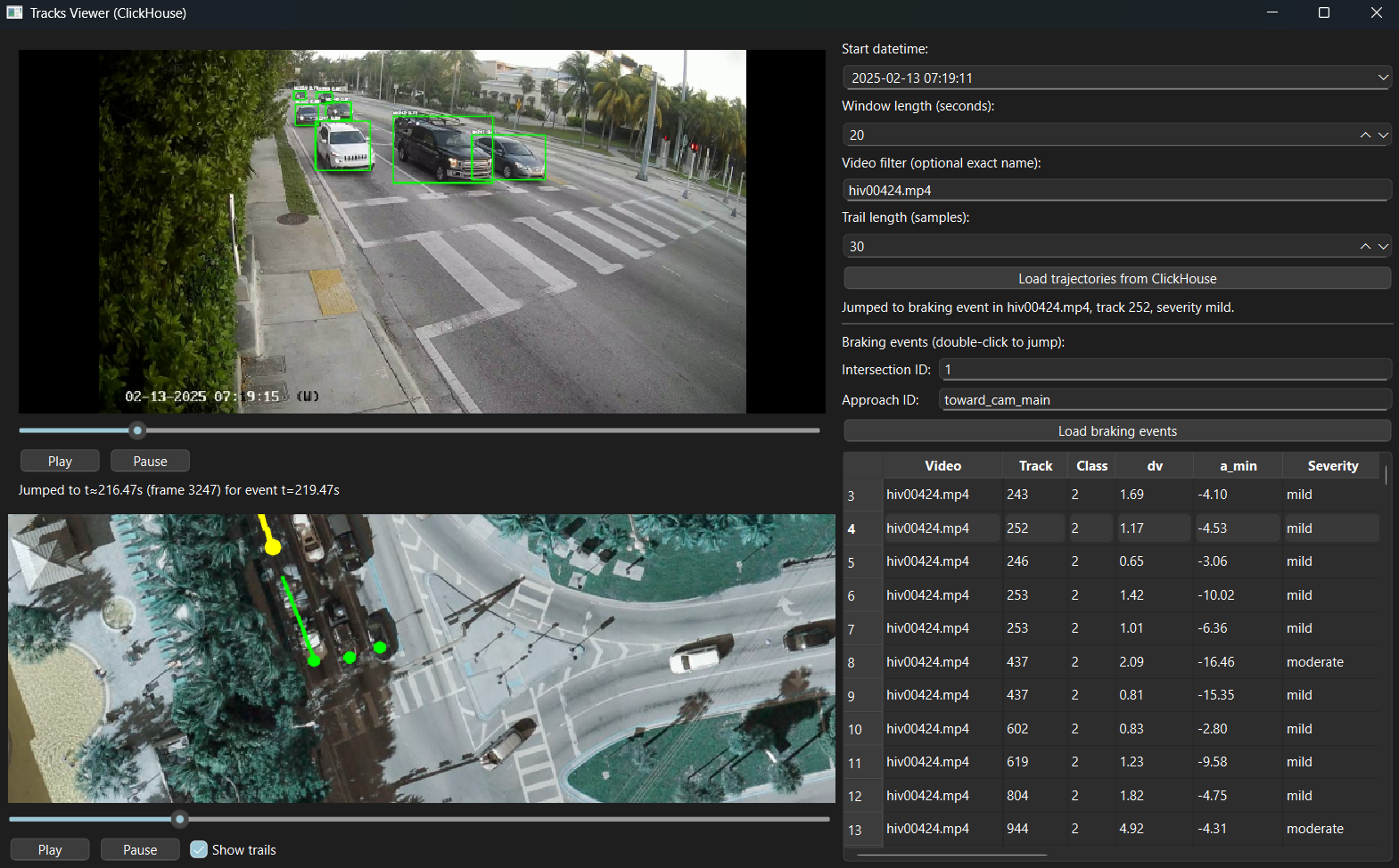}
    \caption{A screenshot of the Tracks Viewer interface.}
    \label{fig:tracksviewer}
\end{figure*}

\subsubsection{Object Detection and Tracking}
\label{sec:objectdet}

The next portion of the pipeline involves ingesting traffic videos to perform object detection and tracking. For the detection model, YOLO11l (by \citeauthor{yolo11_ultralytics}, \citeyear{yolo11_ultralytics})
 was chosen because larger-sized model variants are more capable of tracking many vehicles simultaneously. The YOLO11l model made available by Ultralytics is pretrained on the COCO dataset, which is a large-scale annotated object detection dataset~\citep{lin2015microsoftcococommonobjects}. The pipeline configures the model to an input resolution of 544 $\times$ 960 to preserve the native 1080p traffic camera aspect ratio, aligning the input tensor to the network stride.

Firstly, the timestamp of the video must be determined. Some traffic camera configurations do not always store machine-readable metadata (which entails recording the date and time within a structured text format or database). In the configuration of the case study, the timestamp is present only within the bottom left corner of the video frame, as visible in Figure~\ref{fig:camera_overlay}. Thus, to extract that timestamp, conventional OCR (Optical Character Recognition) was considered initially. However, nonstandard font coloring proved to be difficult to process with OCR. Instead, the application attaches the initial frame of the traffic video inside a query to the Gemma 3-27B-IT vision-language model, which returns the start date and time of the video.

Once the date and time are identified, video processing begins to perform object detection and tracking. Using the homography computed in Equation~\ref{eq:numpymatrix}, YOLO object detections are converted to $\mathbf{H}$ space. Several tracking features are utilized during this process. Firstly, if the camera position changes from video to video, the $\mathbf{H}$ can be dynamically chosen according to the time-of-day or video filename. Secondly, some shortcomings of object tracking, especially within the constraints of the traffic camera, can be managed.

For instance, temporary occlusion of vehicles in far-side lanes can occur, pushing the occluded vehicle's bounding box around despite the object remaining stationary. Furthermore, pretrained YOLO models and trackers may impose jitter on far-away stationary objects. The software utilizes exponential moving average (EMA) smoothing to mitigate the effect of jitter. The generated data are anonymized, protecting road users, as only the trajectory data and no personally identifiable information is stored. The data are stored in a ClickHouse database, which was chosen for its fast data retrieval and ingestion.

The detection/tracking software is publicly available online.\footnote{\url{https://github.com/jpfleischer/TrackerMALT}}

\subsubsection{Braking Metrics Analysis}

To derive braking events, trajectory data is queried and extracted from the ClickHouse database. The following metrics are computed using the extracted data: radial distance from the intersection stop bar, speed and acceleration time series, and denoised deceleration (braking) events. Acceleration is computed as the second time derivative of radial distance to the stop bar. 

It is vital to distinguish object tracker noise from legitimate braking events. Therefore, a parameter for a minimum threshold to indicate deceleration is defined as $a_\text{trigger} = 0.25~\mathrm{m/s^2}$. Using this parameter, several checks are made against candidate events, including:

\begin{enumerate}
    \item The magnitude of the radial acceleration must exceed $a_\text{trigger}$ for a contiguous portion of the event
    \item The robust deceleration, which is defined as the 5th percentile of the radial acceleration time series, must have a magnitude of at least $0.85a_\text{trigger}$
    \item The event must last for at least 0.2 seconds
\end{enumerate}

If a candidate event passes these checks, it is categorized by severity based on the average deceleration $\bar{a}$ and the standard unit for gravity, $g = 9.81~\mathrm{m/s^2}$, across three categories:

\begin{itemize}
    \item \textbf{Mild:}   $0.15g \le |\bar{a}| < 0.25g$
    \item \textbf{Moderate:} $0.25g \le |\bar{a}| < 0.40g$
    \item \textbf{Severe:} $|\bar{a}| \ge 0.40g$
\end{itemize}

The identified braking events can be reviewed using the Tracks Viewer interface of the developed application, which is depicted in Figure~\ref{fig:tracksviewer}. Within the interface, users can review footage and details surrounding events, including timestamps and severity levels. The yellow trajectory in the satellite view represents the pertinent braking vehicle, whereas the green tracks represent surrounding vehicles.



\section{Experimental Results}

This section presents the results of the real-world case study, which demonstrates the pipeline's ability to collect hard brake statistics and their resulting usefulness.

\begin{figure*}[tb]
    \centering
    \includegraphics[width=0.9\linewidth]{{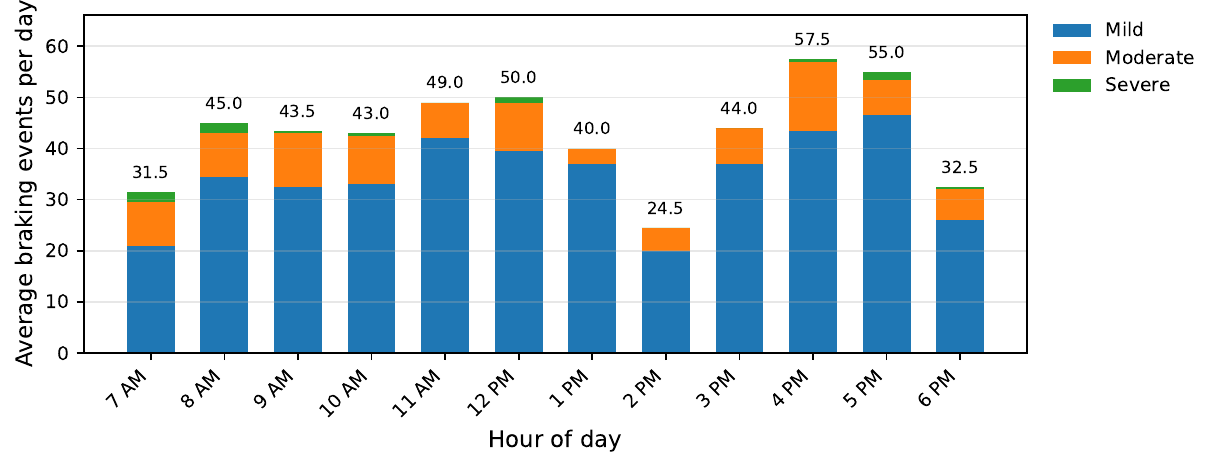}}
    \caption{Average hourly braking event counts by severity (mild, moderate, severe) over 2 days of traffic footage at Overseas Hwy / Roosevelt Blvd.}
    \label{fig:hourlyevents_overseas_roosevelt}
\end{figure*}

\begin{figure*}[tb]
    \centering
    \includegraphics[width=0.9\linewidth]{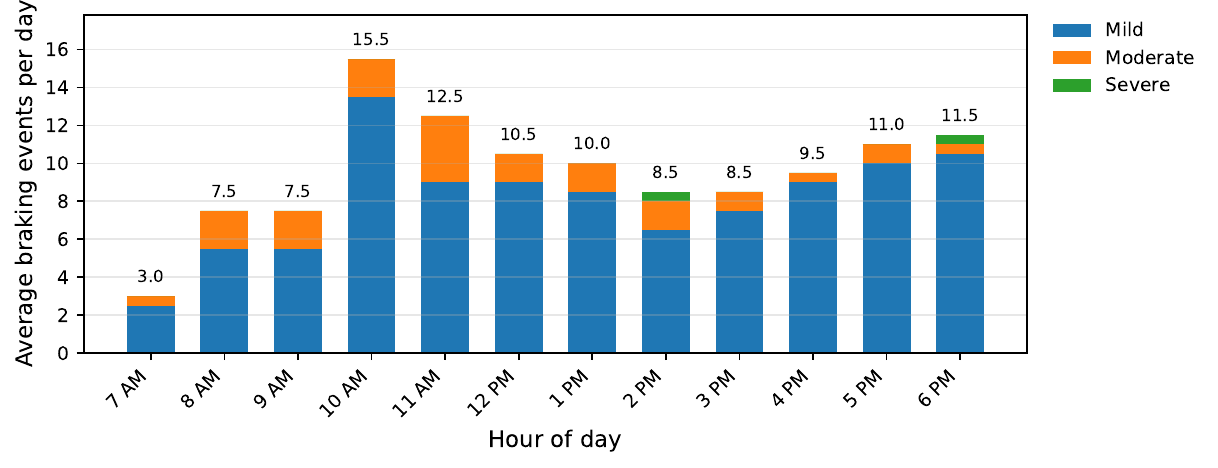}
    \caption{Average hourly braking event counts by severity (mild, moderate, severe) over 2 days of traffic footage at White Street / Truman Avenue.}
    \label{fig:hourlyevents_truman_white}
\end{figure*}

To understand the temporal distribution of braking events during daytime hours, Figure~\ref{fig:hourlyevents_overseas_roosevelt} shows the time-of-day variation in braking statistics estimated based on the data collected by the application. The results in Figure~\ref{fig:hourlyevents_overseas_roosevelt} show three distinct peak periods observed at the Overseas Hwy.\ \& Roosevelt Blvd.\ intersection, specifically the hours of 8 AM, 12 PM, and 4 PM. The number of braking events was high during the AM peak (8:00 AM–10:59 AM), remained relatively high during the midday peak (11:00 AM–1:59 PM), after which it decreased but increased again during the PM peak (3:00 PM–5:59 PM), reflecting the increased traffic demand and more vehicle interactions during these times. The results were somewhat similar at the White St.\ and Truman Ave.\ intersection, with peaks at the 8-9 AM hours, but additional peaks occurring at the hours of 10 AM and 6 PM as shown in Figure~\ref{fig:hourlyevents_truman_white}.

\begin{figure}
    \centering
    \includegraphics[width=0.99\linewidth]{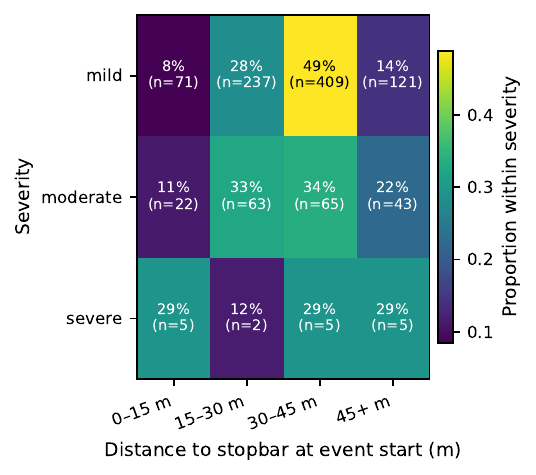}
    \caption{Severity vs distance heatmap (row-normalized proportions) at Overseas \& Roosevelt}
    \label{fig:severitybydistance}
\end{figure}

\begin{figure}
    \centering
    \includegraphics[width=0.99\linewidth]{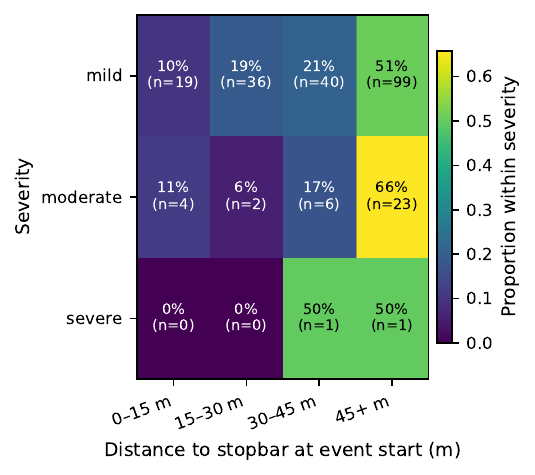}
    \caption{Severity vs distance heatmap (row-normalized proportions) for Truman \& White}
    \label{fig:severitybydistance-truman}
\end{figure}

For further understanding of the braking behavior, the braking events were grouped according to their distance from the intersection stop bar ($r_{start}$) at the beginning of the braking event (0–15 m, 15–30 m, 30–45 m, and greater than 45 m). This allowed for the examination of spatial variations in braking severity as the vehicles approached the intersection. Figure~\ref{fig:severitybydistance} demonstrates that most braking events at Overseas Hwy.\ and Roosevelt Blvd.\ began from 30–45 m upstream of the stop bar across all deceleration severity levels. Mild braking mainly happened farther from the intersection, with 49 percent of mild events occurring 30–45 m from the stop line. This indicates that earlier decelerations are smoother. In contrast, severe braking was consistent at every distance group except for 15–30 m, where it was less likely to occur. At the White St.\ and Truman Ave.\ intersection, a similar pattern was observed in Figure~\ref{fig:severitybydistance-truman}, where most braking events occurred far upstream and were of mild severity (40 and 99 events for 30–45 m and 45+ m upstream of the stop line, respectively). 

The distributions of average braking deceleration by hour of day were analyzed as shown in Tables~\ref{tab:hourly_avg_decel_stats_overseas_roosevelt} and~\ref{tab:hourly_avg_decel_stats_white_truman}. The average braking deceleration was found to have fairly consistent median values throughout the day. However, at the Overseas and Roosevelt intersection, higher extreme deceleration values (90th percentile) occurred during the AM and PM peaks. At the White and Truman intersection, severe braking was most evident during the late morning to afternoon period, despite lower overall braking events.

\begin{figure}[tb]
    \centering
    \includegraphics[width=0.9\linewidth]{{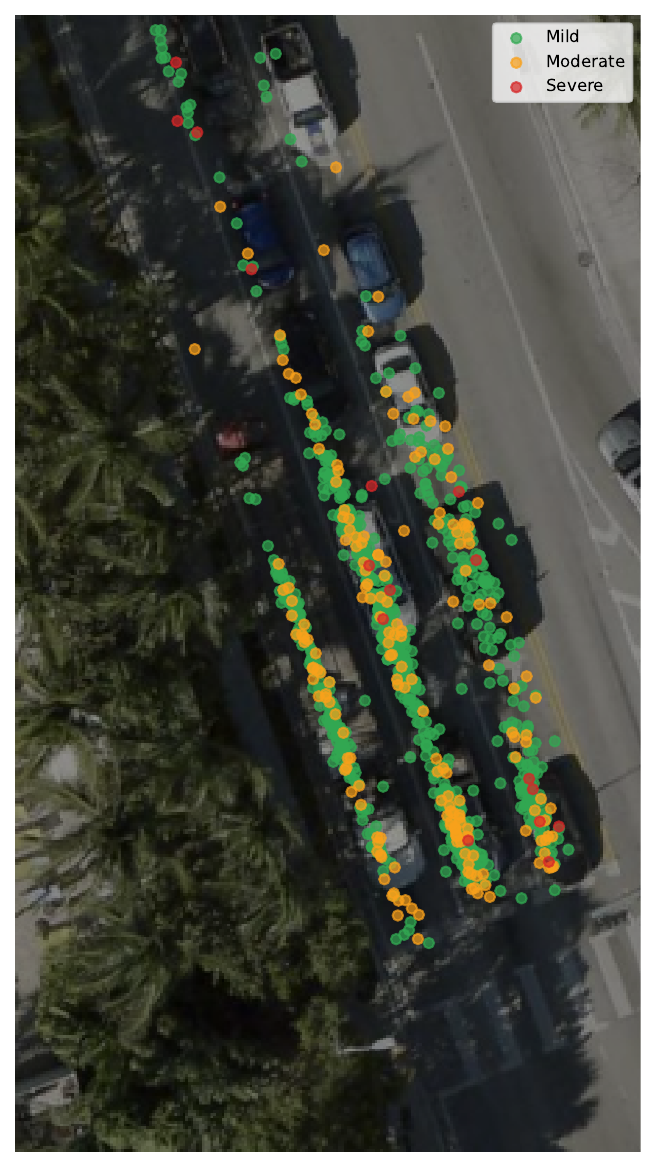}}
    \caption{Average vehicle position over braking event overlaid on the satellite orthophoto at Overseas Hwy / Roosevelt Blvd. Points are colored by severity (mild, moderate, severe).}
    \label{fig:heatmap_scatter_overseas_roosevelt_count}
\end{figure}

\begin{figure*}[tb]
    \centering
    \includegraphics[width=0.9\linewidth]{{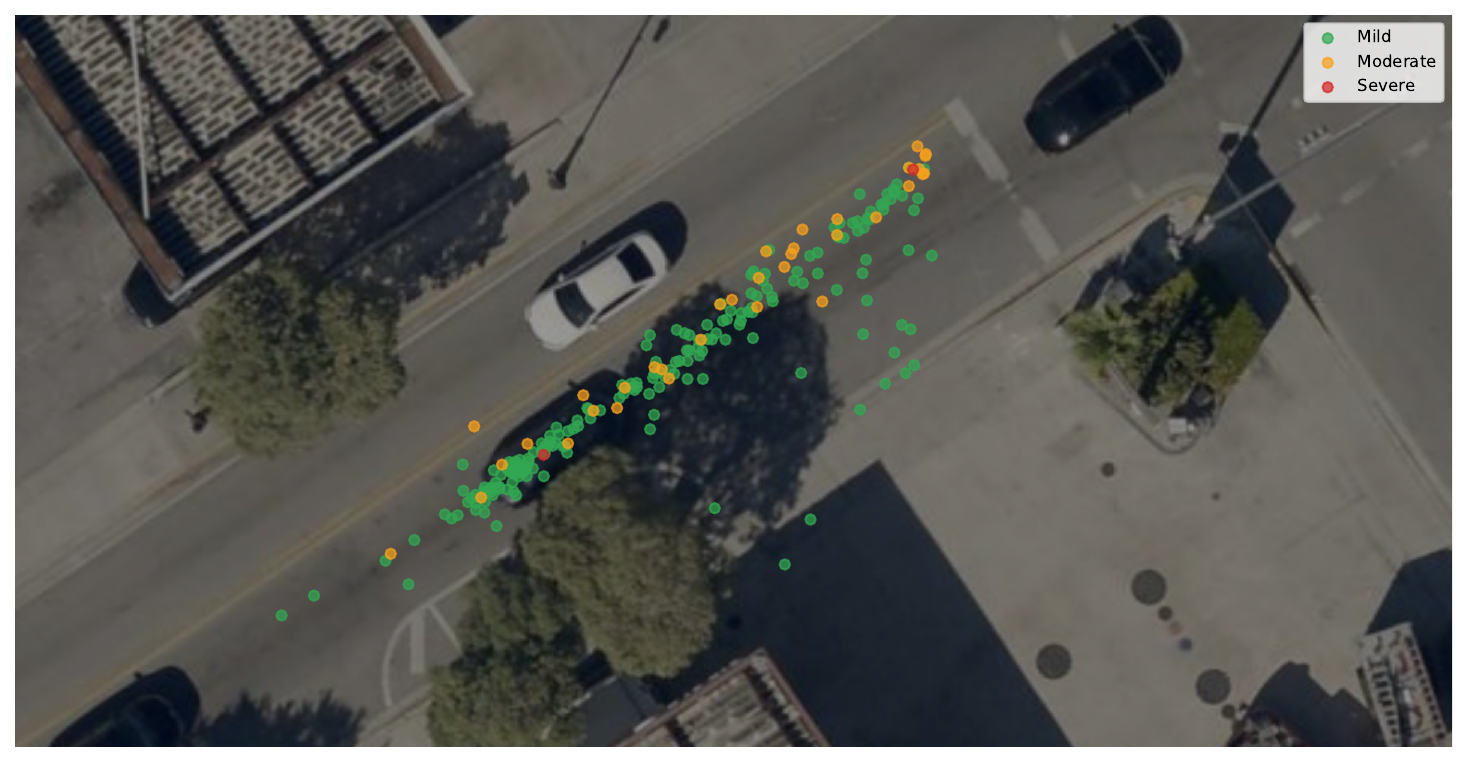}}
    \caption{Average vehicle position over braking event overlaid on orthophotography at White Street / Truman Avenue. Points are colored by severity (mild, moderate, severe).}
    \label{fig:heatmap_scatter_white_truman_count}
\end{figure*}

Figure~\ref{fig:heatmap_scatter_overseas_roosevelt_count} illustrates the average vehicle position over each braking event at the Overseas Hwy.\ and Roosevelt Blvd.\ intersection, while Figure~\ref{fig:heatmap_scatter_white_truman_count} presents the corresponding results for the White St. and Truman Ave. intersection. The result from the Overseas Hwy.\ and Roosevelt Blvd.\ intersection shows that mild and moderate braking events exhibit similar spatial patterns across the upstream approach, consistent with braking start locations. Several severe events occur approximately 15 m. and 30 m. upstream of the stop line in the middle lane and left-turn bay near the stop bar, while no severe braking events are observed in the rightmost lane. For the White St.\ and Truman Ave.\ intersection, braking events are distributed along the upstream approach lane, with mild and moderate events occurring throughout the approach. Several points appear near the adjacent gas station driveway, reflecting braking behavior associated with vehicles attempting to merge into the flow of traffic.

\begin{figure}[tb]
    \centering
    \includegraphics[width=0.9\linewidth]{{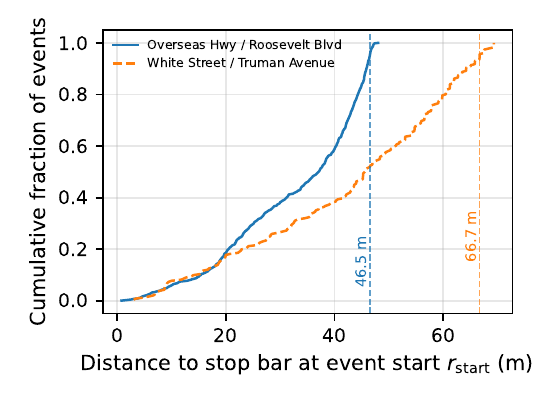}}
    \caption{Empirical cumulative distributions of braking start distance $r_{\text{start}}$ from the stop bar at both study intersections. Vertical dashed lines mark the 95th percentile of $r_{\text{start}}$ for each intersection.}
    \label{fig:rstart_cdf_both_intersections}
\end{figure}

To measure the distance that braking events can be detected upstream, Figure~\ref{fig:rstart_cdf_both_intersections} plots the distance from the stop bar ($r_{start}$) during the beginning of the event for both intersections. Notably, for the Overseas and Roosevelt intersection, the 95th percentile of this distance is 46.5 m, and for the White and Truman intersection, it is 66.7 m. It is vital to note that $r_{start}$ in Figure~\ref{fig:rstart_cdf_both_intersections} depicts the beginning of the braking event instead of the average position over the course of the event (the average positions are plotted in Figures~\ref{fig:heatmap_scatter_overseas_roosevelt_count} and~\ref{fig:heatmap_scatter_white_truman_count}).

The spatial distribution of braking events reflects typical driver behavior on the signalized intersection approaches, where upstream mild and moderate braking corresponds to anticipatory deceleration, while severe braking closer to the stop bar indicates late responses to signal changes or vehicle interactions. The lack of severe events in the rightmost lane suggests lower conflict exposure in that lane. At the White Street and Truman Avenue intersection, braking near the adjacent gas station driveway highlights the impact of access management and merging activity on localized deceleration behavior.

\begin{table*}[tb]
    \centering
    \renewcommand{\arraystretch}{1.15}
    \caption{Hourly average braking deceleration statistics at Overseas Hwy / Roosevelt Blvd. Shown are counts and summary measures of $\bar{a}$ (m/s$^2$) for braking events between 7:00 AM and 7:00 PM. Quantiles shown are minimum, 25th, median, 75th, and 90th percentile.}
    \label{tab:hourly_avg_decel_stats_overseas_roosevelt}
    \begin{tabular}{c|rrrrrrr}
        \hline
        Hour & $n$ & Mean & Min & 25th & Med. & 75th & 90th \\
        \hline
        7 AM & 60 & 3.58 & 1.56 & 2.42 & 2.87 & 3.74 & 5.66 \\
        8 AM & 82 & 3.48 & 1.54 & 2.21 & 2.95 & 3.91 & 6.35 \\
        9 AM & 82 & 3.26 & 1.50 & 2.40 & 2.91 & 3.91 & 4.84 \\
        10 AM & 81 & 3.33 & 1.60 & 2.21 & 2.79 & 4.13 & 5.36 \\
        11 AM & 92 & 3.32 & 1.53 & 2.25 & 2.83 & 3.92 & 5.02 \\
        12 PM & 89 & 3.24 & 1.60 & 2.32 & 2.80 & 3.78 & 5.27 \\
        1 PM & 74 & 3.00 & 1.50 & 2.10 & 2.58 & 3.44 & 4.10 \\
        2 PM & 44 & 3.56 & 1.71 & 2.57 & 2.95 & 3.91 & 5.11 \\
        3 PM & 83 & 3.16 & 1.61 & 2.08 & 2.71 & 3.48 & 5.09 \\
        4 PM & 108 & 3.37 & 1.53 & 2.26 & 2.79 & 3.95 & 4.89 \\
        5 PM & 99 & 2.94 & 1.48 & 2.24 & 2.62 & 3.41 & 4.20 \\
        6 PM & 61 & 3.31 & 1.49 & 2.15 & 2.69 & 4.03 & 5.38 \\
        \hline
    \end{tabular}
\end{table*}

\begin{table*}[tb]
    \centering
    \renewcommand{\arraystretch}{1.15}
    \caption{Hourly average braking deceleration statistics at White Street / Truman Avenue. Shown are counts and summary measures of $\bar{a}$ (m/s$^2$) for braking events between 7:00 AM and 7:00 PM. Quantiles shown are minimum, 25th, median, 75th, and 90th percentile.}
    \label{tab:hourly_avg_decel_stats_white_truman}
    \begin{tabular}{c|rrrrrrr}
        \hline
        Hour & $n$ & Mean & Min & 25th & Med. & 75th & 90th \\
        \hline
        7 AM & 6 & 2.48 & 1.67 & 2.23 & 2.23 & 2.79 & 3.26 \\
        8 AM & 15 & 3.18 & 1.93 & 2.43 & 2.72 & 3.39 & 5.25 \\
        9 AM & 15 & 3.79 & 1.92 & 2.55 & 2.91 & 3.49 & 4.50 \\
        10 AM & 31 & 3.85 & 1.68 & 2.62 & 3.29 & 4.92 & 6.31 \\
        11 AM & 25 & 3.83 & 1.53 & 2.50 & 3.46 & 4.32 & 6.07 \\
        12 PM & 21 & 3.60 & 1.65 & 2.35 & 2.90 & 4.14 & 4.78 \\
        1 PM & 20 & 3.51 & 1.84 & 2.46 & 2.93 & 3.85 & 6.29 \\
        2 PM & 17 & 3.93 & 1.53 & 2.96 & 3.55 & 4.04 & 6.76 \\
        3 PM & 17 & 4.22 & 1.93 & 2.26 & 3.12 & 5.52 & 6.58 \\
        4 PM & 19 & 3.83 & 2.02 & 2.35 & 3.51 & 4.46 & 6.58 \\
        5 PM & 22 & 2.81 & 1.62 & 2.17 & 2.68 & 3.01 & 3.98 \\
        6 PM & 23 & 3.26 & 1.54 & 2.07 & 3.22 & 3.51 & 3.89 \\
        \hline
    \end{tabular}
\end{table*}

\section{Related Works}

Prior research has demonstrated the feasibility of extracting vehicle trajectories and safety-related metrics from visual sensing systems. High-accuracy localization has been achieved by matching traffic camera imagery with high-definition drone-derived maps \citep{He2022}; however, such approaches require specialized equipment and operational overhead, limiting their scalability. Alternatively, several studies have relied on homography transformations to project oblique camera views into a top-down perspective for estimating vehicle speed, lane changes, and trajectories \citep{10026962}. While effective, homography-based methods typically lack an externally defined metric scale and often require precise camera calibration.

To overcome these limitations, recent work has incorporated satellite imagery to provide spatial reference frames for monocular camera systems. For example, \citet{mejia2021vehicle} utilized Google Maps imagery for camera calibration to estimate vehicle speed. Building upon this concept, the present study aligns traffic camera views with georeferenced orthoimagery retrieved programmatically via an ArcGIS REST service, ensuring metric reliability and enabling direct integration with Geographic Information System (GIS) layers. Robust estimation techniques, such as MAGSAC++ \citep{barath2019magsacfastreliableaccurate}, are employed to improve alignment accuracy in the presence of feature outliers. Recent advances in deep learning-based object detection further support scalable camera-based analysis; in particular, pretrained YOLO models provide accurate real-time vehicle detection without the need for custom training datasets \citep{yolo11_ultralytics}.

Hard-braking behavior has been widely recognized as a meaningful surrogate safety measure, especially for rear-end collision risk at signalized intersections. Empirical studies comparing multi-year crash records with hard-braking event data have identified strong correlations between braking frequency and crash occurrence, particularly for braking events occurring upstream of stop bars \citep{article}. Other approaches estimate braking behavior using Vehicle-to-Everything (V2X) data increasingly collected and monetized by Original Equipment Manufactures (OEMs), where vehicle deceleration profiles are derived from connected vehicle communications \citep{Sengupta2023}. While such methods provide high-fidelity kinematic data, they are inherently constrained by connected-vehicle penetration rates. Simulation-based evaluations of Connected and Automated Vehicle (CAV) systems further suggest that substantial safety benefits from advisory systems require high market penetration \citep{doi:10.1177/0361198118797831}.

In contrast to sensor-intensive or connectivity-dependent approaches, this study extends prior work by enabling braking behavior measurement for all vehicles within the field of view of existing traffic cameras. By combining georeferenced orthoimagery, robust camera-to-map alignment, and modern deep learning-based detection, the proposed software tool provides a scalable and cost-effective solution for measuring braking behavior and other safety-critical traffic metrics without requiring specialized hardware or connected-vehicle infrastructure.

\section{Conclusions}
\label{sec:conclusion}


This paper demonstrates the effectiveness and usefulness of a machine learning-based machine vision technique in the collection of braking statistics. The analysis of the data collected for the case study showed the ability of the application to identify clear temporal and spatial patterns in braking behavior at the study intersections. The collected data indicated that the braking frequency varied by time of day. These peaks correspond to increased traffic and vehicle interactions. This information allows transportation agencies to identify locations and periods in the day that have operational and safety issues that need to be addressed and prioritized when developing improvement alternatives. The spatial analysis confirmed that most braking events started before the stop bar. Mild braking mostly happened farther from the intersection, suggesting smoother deceleration. In contrast, severe braking appeared consistent at every distance but less likely at 15–30 m. Severe braking depicted sudden stops, which may correspond to higher crashes, particularly rear-end crashes. The relationships between the braking characteristics and crash statistics will be investigated in a future study. While the average braking deceleration stayed fairly stable throughout the day, peak braking deceleration showed varying median values and more variability by time-of-day. This reveals that sudden and critical braking events need to be investigated to further determine their relationship to crash occurrence. These findings emphasize the need to consider both time patterns and distance to the stop bar when assessing braking behavior and intersection safety.

Future research will focus on improving trajectory robustness under occlusions and adverse weather conditions, extending the framework to multi-camera fusion for corridor- and network-level analysis, and integrating additional surrogate safety measures such as time-to-collision and post-encroachment time. Further work will also explore real-time deployment and validation against ground-truth connected vehicle to support operational traffic safety applications.




\bibliographystyle{apalike}
\begin{small}
    \bibliography{references}
\end{small}

\end{document}